\documentclass[10pt,letterpaper]{article} 
\usepackage{titlesec,float,makeidx,multirow,multicol,graphicx,hyperref,amsmath,amssymb,fancyhdr,changepage,etoolbox,times,enumitem,textcomp,indentfirst,subcaption,caption,setspace}

\usepackage[dvipsnames,svgnames,table]{xcolor}
\usepackage{txfonts}
\usepackage{mathdots}
\usepackage[paperwidth=8.5in,paperheight=11in,top=1in,right=0.5in,bottom=1in,left=0.5in]{geometry}
\usepackage[utf8]{inputenc}
\usepackage[sort&compress,numbers]{natbib} 
\fancypagestyle{firstpage}{
\fancyhead[L,C,R]{}
\fancyfoot[R]{ }
\fancyfoot[C]{\thepage}
\fancyfoot[L]{}

}

\RequirePackage{fix-cm}
\newcommand{\size}[2]{{\fontsize{#1}{0}\selectfont#2}}

\titleformat{\section}
  {\normalfont\fontfamily{phv}\bfseries}{\thesection}{10pt}{\MakeUppercase}
\renewcommand*{\thesection}{\fontfamily{phv}\selectfont\textbf{\arabic{section}.}}
\titlespacing{\section}{0pt}{10pt}{0pt}
\titleformat{\subsection}
  {\normalfont\fontfamily{phv}\bfseries}{\thesubsection}{3pt}{}
\renewcommand*{\thesubsection}{\fontfamily{phv}\selectfont\textbf{\arabic{section}.\arabic{subsection}}}
\titlespacing{\subsection}{0pt}{10pt}{0pt}

\DeclareCaptionFormat{upper}{#1#2\uppercase{#3}\par}
\newcommand{\startsquarepar}{%
    \par\begingroup \parfillskip 0pt \relax}
\newcommand{\stopsquarepar}{%
    \par\endgroup}

\hypersetup{
	colorlinks=true,
	linkcolor=black,
	filecolor=blue,      
	urlcolor=blue,
	citecolor=black,
}
\usepackage{nomencl}
\usepackage{bm}
\usepackage{xstring}
\usepackage{xpatch}
\patchcmd{\thenomenclature}
  {\leftmargin\labelwidth}
  {\leftmargin\labelwidth\itemindent 0.25in }
  {}{}

\makenomenclature

\begin{document}
\newgeometry{top=0.5in,right=0.5in,bottom=1in,left=0.5in}
\begin{flushright}
\fontfamily{phv}\selectfont{
\textbf{Conference Paper}\\
\textbf{2022}\\
\textbf{ }\\
\textbf{ }\\[45pt]
\textbf{\size{18}{ }}\\[38pt]
}
\end{flushright}
\begin{center}
    \fontfamily{phv}\selectfont{\size{11}{\textbf{Dynamic Transition From Regular to Mach Reflection Over a Moving Wedge\\[20pt]}}}
\end{center}


    \begin{flushright}
        \fontfamily{phv}\selectfont{\textbf{Lubna Margha$^{1,3}$, Ahmed A. Hamada$^{1}$, Doyle D. Knight$^{2}$, Ahmed Eltaweel$^{4}$} \newline
        $^1$Department of Ocean Engineering, Texas A$\&$M University, College Station, TX, 77843, USA \newline $^2$Department of Mechanical and Aerospace Engineering, Rutgers - The State University of New Jersey, Piscataway, NJ, 08854, USA \newline
        $^3$Department of Aeronautical and Aerospace Engineering, Cairo University, Giza 12613, Egypt \newline 
         $^4$Aerospace Engineering Program, University of Science and Technology - Zewail City, Giza, 12578, Egypt \newline }
    \end{flushright}
    
\begin{multicols*}{2}
\section*{Abstract}
\textit{The transition between the Regular Reflection (RR) and Mach Reflection (MR) phenomenon impacts the design of the supersonic and hypersonic air-breathing vehicles. The aim of this paper is to numerically investigate the dynamic transition from RR to MR of unsteady supersonic flow over a two-dimensional wedge, whose trailing edge moves along the $x$-direction upstream with a velocity, $V(t)$ at a free-stream Mach number of $M_{\infty}=3$. The simulation is conducted using the unsteady compressible inviscid flow solver, which is implemented in OpenFOAM\textsuperscript{\textregistered}, the open-source CFD tool. Further, the wedge motion is applied by moving the mesh boundary, performing the Arbitrary Lagrangian-Eulerian (ALE) technique. In addition, the sonic and detachment criteria are used to define the dynamic transition from RR to MR during the increase of the wedge angle. Different reduced frequencies, $\kappa$, in the range of $[0.1-2]$ for the moving wedge are applied to study the lag in the dynamic transition from the steady-state condition. The results show that the critical value of $\kappa=0.4$ distinguishes between the rapid and gradual lag in the transition from RR to MR. In addition, the transition from RR to MR occurs above the Dual Solution Domain (DSD), since the shock is curved downstream during the rapid motion of the wedge.}

Keywords: Regular reflection; Mach reflection; Moving wedge; Dynamic shock waves; Supersonic flow; Dual solution domain.
\mbox{}
\nomenclature[B]{$\kappa$}{Reduced frequency}
\nomenclature[A]{$L(t)$}{Wedge stream-wise length}
\nomenclature[A]{$w(t)$}{Wedge chord}
\nomenclature[A]{$\tau$}{Non-dimensional time}
\nomenclature[A]{$\theta(t)$}{Wedge angle}
\nomenclature[A]{$h$}{Wedge height}
\nomenclature[A]{$M_{\infty}$}{Free-stream Mach number}
\nomenclature[A]{$M_t(t)$}{Trailing-edge Mach number}
\nomenclature[A]{$V_t(t)$}{Trailing-edge velocity}
\nomenclature[A]{$\mathrm{MS}$}{Mach stem height}
\nomenclature[A]{$H$}{Half height of computational inflow boundary}
\nomenclature[B]{$\beta_{p}$}{Incident wave angle at reflection$/$triple point}
\nomenclature[B]{$\beta_{tang}$}{Tangent wave angle at wedge's apex}
\nomenclature[B]{$\beta_{t}$}{Transition wave angle}
\printnomenclature[0.68in]
\section{Introduction}
\begin{figure}[H]
	\centering
	\begin{subfigure}{0.49\linewidth}
        \includegraphics[width=0.99\linewidth]{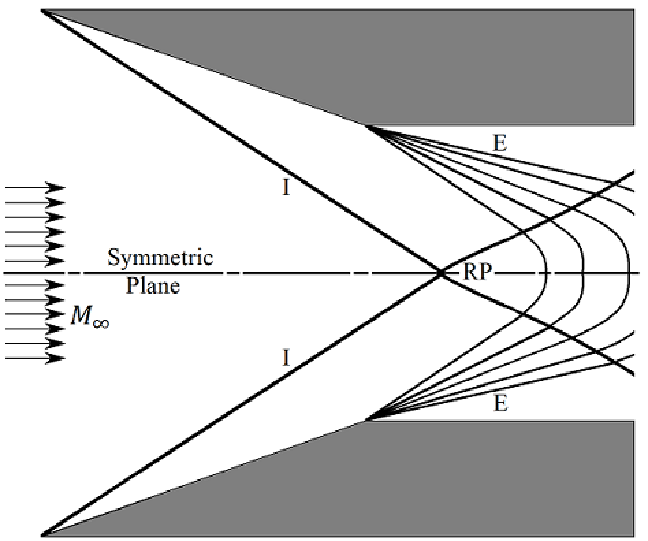}
        \caption{RR configuration}\label{fig:fig1a}
	\end{subfigure}
	\begin{subfigure}{0.49\linewidth}
        \includegraphics[width=0.99\linewidth]{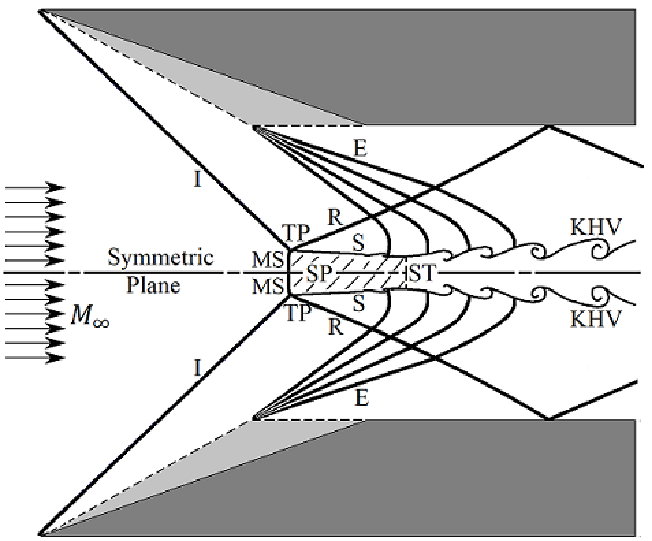}
        \caption{MR configuration}\label{fig:fig1b}
	\end{subfigure}
\caption{ The shock structure of the RR and MR configurations, respectively.}\label{fig:fig1}
\end{figure}

Predicting the shock reflections and the shock wave interactions are very crucial in the design and operation phases of many engineering applications, such as shock-wave focusing, protection against blasts and detonations, and supersonic and hypersonic vehicles. For instance, a series of shock wave interactions are generated in the scramjet inlet, using tilted wedges, to decelerate the flow and achieve efficacious combustion. A proper comprehension of dynamic shock wave interactions over simple moving geometries, such as a wedge, will give insights into the limitations of the moving supersonic intake for an efficient operation.
\startsquarepar \indent When a supersonic flow impinges a wedge, an incident shock is generated. Then, it reflects on the mid-plane of symmetry, creating a second shock. Figure \ref{fig:fig1} shows that the reflection of the incident shock on the mid-plane will follow one of two configurations, Regular Reflection (RR) or Mach Reflection (MR), which depends on the free-stream Mach number, $M_{\infty}$, and the incident shock angle, $\beta$, \cite{hornung1986regular,ben2007shock}. The structure of RR is formed of two shock waves, the incident shock wave (I), and the reflected shock wave (R), as shown in Figure \ref{fig:fig1a}. They gather on the reflecting surface at a point, called the reflection point (RP). The reflected shock deforms slightly from a straight line due to the interference with the Expansion fans (E), generated at the trailing edge of the wedge. On the other hand, the structure of the MR includes three shock waves, the incident shock wave (I),\stopsquarepar
\end{multicols*}
\restoregeometry
\clearpage
\begin{multicols*}{2}
\noindent the reflected shock wave (R), and the Mach stem (MS), which all meet at the triple point (TP) and a slip line (S) appears, as shown in Figure \ref{fig:fig1b}. The slip line is formed due to the difference in the flow parameters behind the reflected shock and the Mach stem. The subsonic region behind the Mach stem bounded by the slip lines and the sonic throat (ST), is called the subsonic pocket (SP). Again, expansion fans (E), interfere with the reflected shock and bend it. The weak waves, that propagate behind the reflected shock from expansion waves, reach the slip line causing the generation of Kelvin-Helmholtz vortices (KHV). The transition from RR to MR over a wedge occurs when the wedge angle is large enough, that the reflected shock is no longer able to turn the flow parallel to the mid-plane. Thus, a normal shock is created and MR happens.
\indent John von Neumann \cite{von1943oblique} is the first one who introduced two different criteria for the transition between RR and MR for symmetrical reflection, known as the detachment criterion and the von Neumann (or mechanical equilibrium) criterion. The detachment criterion is used to determine the transition from RR to MR and denotes the maximum shock deflection angle $\beta_{D}$ that a RR configuration is theoretically possible. At a higher angle, the reflection point is forced to detach from the mid-plane forming the Mach stem. The von Neumann criterion denotes the theoretical limit of the shock deflection angle $\beta_N$ for the MR configuration. Subsequently, the length scale criterion is proposed by Hornung et al. \cite{hornung1979transition}. This criterion assumes the MR occurs at a certain length of the Mach stem. Moreover, the sonic criterion is also used to define the transition from RR to MR, when the flow beyond the reflected shock becomes sonic which assures the beginning of the Mach stem. Thus, the detachment criterion and the sonic criterion are very close. Furthermore, Both the RR and MR are possible solutions to exist in a range of wave angles ($\beta_N < \beta < \beta_D$) at relatively high Mach numbers ($M_{\infty}>2.2$ for a perfect gas with the specific heat ratio $\gamma = 1.4$), which called Dual Solution Domain (DSD) \cite{hornung1982transition}. The supposition of hysteresis within the DSD in steady high-speed flows was first assumed by Hornung et al. \cite{hornung1979transition}. Then, they tried to confirm this hypothesis experimentally \cite{hornung1982transition}, but they observed no hysteresis because of the disturbances in the wind tunnel. After that, Chpoun et al. \cite{chpoun1995reconsideration} well studied the hysteresis phenomenon experimentally. In addition, Ivanov et al. used different numerical methods, such as Direct Simulation Monte Carlo \cite{ivanov1995hysteresis}, and Euler approach \cite{ivanov2001transition}, and also used the kinetic and continuum approaches \cite{ivanov1998numerical}, to carefully investigate the hysteresis.\\
\indent The transition between RR and MR within the DSD can be achieved using the pulsed energy deposition \cite{yan2003effect,khotyanovsky2004parallel}, or the movement of the wedge \cite{felthun2002transition}. Felthun and Skews \cite{felthun2004dynamic} simulated the dynamic transition between RR and MR over a rotating wedge around its leading edge at $M_{\infty}=3$ and various rates. They found that the nature of the transition is dependent on the rate of the wedge rotation. At a very low trailing-edge Mach number ($M_t=0.001$), the transition occurs near the theoretical limit. On the other hand, when the wedge angle changes at higher rates, the transition angles are beyond the DSD. Further, the lag between the dynamic transition angle, $\beta_{t}$, from RR to MR and the transition wave angle in the static case ($\beta_{t_{SC}}=39.5^o$) increases rapidly at relatively low $M_t$ to an asymptotic value around $42^o$ at $M_t>0.05$. Later on, Naidoo and Skews \cite{naidoo2011dynamic} continued the work using a rapidly rotating wedge and studied the transition between RR and MR in the weak and strong-reflection ranges. Their experimental and numerical results showed that the transition from RR to MR happens beyond the steady-state detachment criterion, and the transition from MR to RR occurs below the von Neumann theoretical limit. Furthermore, Ivanov et al. \cite{ivanov2001transition} conducted numerical and experimental investigations over a rotating wedge around its trailing edge in a supersonic flow with $M_{\infty}=4$. The numerical simulations were used to study the hysteresis. In addition, their experiments showed that the transition between RR and MR occurs around the von Neumann angle.\\
\indent Numerical simulations were conducted in the present work to investigate the dynamic transition from RR to MR using a two-dimensional wedge at a free-stream Mach number $M_{\infty}=3$. The trailing edge of the wedge moves horizontally upstream (the wedge height is kept constant) with a constant reduced frequency, $\kappa$. Further, different values of $\kappa$ were tested to study the lag in the transition, and the development of the Mach stem height.


\section{Physical Model}
\subsection{Model Description}
\begin{figure}[H]
    \centering
    \includegraphics[width=0.99\linewidth]{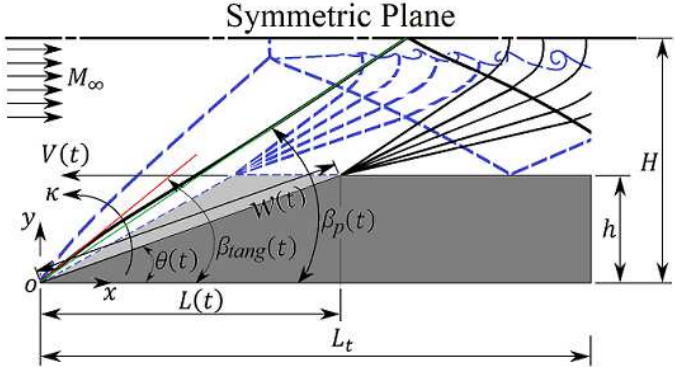}
    \caption{ The flow configuration of the horizontal motion of two symmetrical wedges in a supersonic flow.}\label{fig:fig2}
\end{figure}

The flow configuration of two symmetrical wedges in a supersonic flow with a free-stream Mach number, $M_\infty= 3$, is shown in Figure \ref{fig:fig2}. The height of the computational inflow boundary is $2H$, the height of each wedge is $h$, the initial wedge chord is $w(0)=1$, the stream-wise length of the wedge is $L(t)$, the length of the wedge plus flat plate is $L_t$, and the time-dependent wedge angle is $\theta(t)$. The trailing edge of the wedge moves in the $x$-direction with velocity, $V (t)$, from its initial location. Thus, the wedge angle, $\theta(t)$, and wedge length, $L(t)$, change during the motion, while $H$, $h$ and $L_t$ remain constant. The flow is assumed two-dimensional and inviscid, and the gas is calorically and thermally perfect. When the supersonic flow incident on a wedge, the wedge angle generates an oblique shock wave that is reflected on the plane of symmetry. According to the angle of the wedge, there are two scenarios for the reflected shock; either a RR configuration or a MR configuration is formed. The dynamic transition from RR to MR during the increase of $\theta (t)$ from $19^{\circ}$ to $34^{\circ}$ is studied for different constant values of $\kappa$ at a range of $[0.1,2]$ with step $0.1$. Due to the lag in the shock system, there are two wave angles that can be measured; wave angle at reflection$/$triple point, $\beta_{p}$, and tangent wave angle at wedge’s apex, $\beta_{tang}$. Table \ref{table:1} shows the values of these parameters.
\begin{table}[H]
\centering
\caption{System properties and parameters.}
\begin{tabular}{c c} 
 \hline \rule{0mm}{2.5ex}
 Initial wedge's chord, $w(0)$                 & $1 m$             \\[0.5ex]
 \hline
 Initial wedge angle, $\theta(0)$              & $19^{\circ}$      \\[0.5ex]
 Final wedge angle, $\theta(t_f)$              & $34^{\circ}$      \\[0.5ex]
 Wedge height to half domain height, $\frac{h}{H}$ & $0.3617$      \\[0.5ex]
 Initial wedge length to half domain height, $\frac{L(0)}{H}$ & $1.0506$\\[0.5ex]
 Total wedge length to half domain height, $\frac{L_t}{H}$ & $2$   \\[0.5ex]
 Free-stream Mach number, $M_{\infty}$         & $3$               \\[0.5ex]
 Reduce frequency, $\kappa$                    & $0.1:0.1:2$       \\[0.5ex]
 \hline
\end{tabular}
\label{table:1}
\end{table}

\subsection{Governing Equations}
Two-dimensional unsteady compressible Euler equations are used to model the supersonic flow over a wedge and are expressed in the conservative form as:	
\begin{equation}
	\frac{\partial Q}{\partial t}+\frac{\partial F}{\partial x}+\frac{\partial G}{\partial y}=0,
\end{equation}
where
\begin{equation}
Q= \begin{bmatrix}
    \rho \\
    \rho u\\
    \rho v\\
    \rho e
\end{bmatrix}, \quad F= \begin{bmatrix}
    \rho u \\
    \rho u^2+p\\
    \rho u v\\
    u (\rho e+p)
    \end{bmatrix}, \quad G= \begin{bmatrix}
    \rho v \\
    \rho u v\\
    \rho v^2+p\\
    v (\rho e+p)
    \end{bmatrix}
\end{equation}
The static pressure is obtained from
\begin{equation}
	p=(\gamma-1) \left( \rho e - \rho \frac{u^2+v^2}{2}\right)
\end{equation}
where $u$ and $v$ are the velocity components in the Cartesian coordinates $x$ and $y$, respectively, $\rho$, $p$ and $e$ are the density, the pressure, and the internal energy of the flow field, respectively, and $\gamma$ is the specific heat ratio of air, which equals $1.4$.

\subsection{Equations of Motion}
The trailing edge of the wedge moves horizontally with velocity $V(t)$ from its initial location with a constant wedge angular velocity, $\omega=d\theta/dt$ ($sec^{-1}$). Because the wedge height, $h$, is kept constant, the increase in the wedge angle, $\theta(t)$, changes only the wedge stream-wise length, $L(t)$, and they are expressed as:
\begin{subequations}
\begin{equation}
	L(t)= h\ cot\left(\theta(t)\right)
\end{equation}
\begin{equation}
	\theta(t)=tan^{-1} \left(\frac{h}{L(0)}\right)+ \omega t
\end{equation}
\end{subequations}
where $L(0)$ is the initial wedge stream-wise length at the starting time, $t=0$, of the motion.\\
\indent Further, the dimensional velocity of the trailing edge of the wedge is expressed as:
\begin{equation}
	V_t(t)=\omega\ h \sqrt{1+cot^{2}\left(\theta(t)\right)}
\end{equation}
\indent In addition, the wedge angular velocity, $\omega$, and the time of the motion, $t$, are normalized using the free-stream velocity and the initial wedge stream-wise length at time $t=0$, $L(0)$, to introduce the non-dimensional frequency, $\kappa$, and the non-dimensional time, $\tau$, which are defined as:
\begin{subequations}
\begin{equation}
	\kappa=\frac{\omega\ L(0)}{U_\infty}
\end{equation} 
\begin{equation}
	\tau=\frac{t\ U_\infty}{L(0)} 
\end{equation} 
\end{subequations}
\indent Therefore, the trailing edge Mach number can be written as:
\begin{equation}
	M_t(\tau)= \kappa\ M_{\infty}\ tan\left(\theta(0)\right) \sqrt{1+cot^{2}\left(\theta(\tau)\right)}
\end{equation}
where $\theta(\tau)$, the wedge angle as a function of non-dimensional time, is defined as: 
\begin{equation}
	\theta(\tau) = \theta(0) + \kappa \tau
\end{equation}
where $\theta(0)$ is the initial wedge angle at the starting non-dimensional time, $\tau=0$, of the motion.
\section{Computational Model}
\subsection{Numerical Implementation}
\textit{rhoCentralDyMFoam} is a transient, density-based compressible flow solver with support for a mesh-motion, that is implemented in OpenFOAM\textsuperscript{\textregistered}$-$v2006. The solver's technique combines the semi-discrete and upwind-central non-staggered schemes of Kurganov and Tadmor \cite{kurganov2000new,kurganov2001semidiscrete}. The previously mentioned schemes are implemented in the solver in order to avoid the use of Riemann solvers of characteristic decomposition \cite{greenshields2010implementation}. In addition, \textit{rhoCentralDyMFoam} depends on an operator-splitting method to solve both momentum and energy equations. Further, explicit predictor equations are implemented for the convection of the conserved variables, whereas implicit corrector equations are used for the diffusion of the primitive variables \cite{greenshields2010implementation}. The van Leer limiter, conducted in the \textit{rhoCentralDyMFoam}, efficiently balances the performance of the solution through shock capture, oscillations-free fields, and computational cost \cite{marcantoni2012high}.

\subsection{Computational Domain}
\begin{figure}[H]
    \centering
    \includegraphics[width=0.99\linewidth]{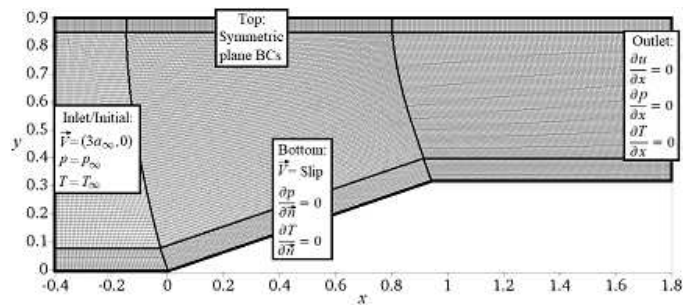}
    \caption{Schematic of the computational domain, the boundary, and initial conditions, for mesh 1}\label{fig:fig3}
\end{figure}

Since the problem is symmetric in the geometry and the flow behavior, only half of the domain is computed. The computational domain is body-fitted for the wedge with an initial wedge angle of $19^{\circ}$, as shown in Figure \ref{fig:fig3}. The streamwise length and the transverse length of the domain are kept constant during the simulations of $2.2w(0)$ and $0.9w(0)$, respectively, where $w(0)$ is the initial wedge chord length at $t = 0$. The origin is placed at the leading edge point of the wedge, and the total length of the wedge, $L_t$, is $1.8w(0)$. The domain is bounded by four boundaries: inlet, outlet, bottom, and top. The flow enters the domain with supersonic Mach number, $M_\infty = 3$, and free-stream pressure and temperature at sea level. At the outlet boundary, a zero gradient boundary condition is applied to all of the flow variables, in order to set the outlet field to the internal field value. The bottom boundary condition (the wedge and the flat plate in front of the wedge) has a slip velocity boundary condition with a zero gradient boundary condition for the pressure and temperature. The symmetric plane boundary conditions are applied at the top surface.

\subsection{Grid Generation}
\begin{figure}[H]
    \centering
    \includegraphics[width=0.93\linewidth]{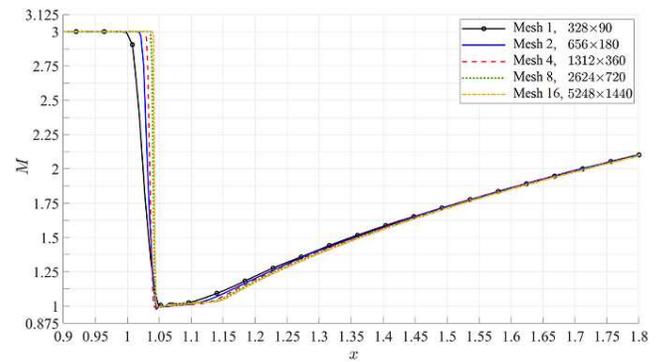}
    \caption{Independent mesh study: The Mach number variation along the top symmetry plane at $y=0.9$, and at the sonic criterion of transition, using different grid sizes.}\label{fig:fig4}
\end{figure}

In this study, the delay in the dynamic transition from the static case value is mainly caused by two reasons; the mesh resolution and the rate of motion of the wedge trailing edge. the moving rate will be discussed further in the results section. Hence, the mesh-independent test is very critical for more accurate results. A 2D-ordered grid of quadrilateral elements was used to discretize the computational domain. The domain was divided into 9 blocks as shown in Figure \ref{fig:fig3}. Orthogonal grids were generated in the blocks near the wedge surface and the mid-plane of symmetry, to reduce the computational noise on the results. Moreover, the middle blocks' edges were curved to improve the grid orthogonality and maintain the same aspect ratio in that region. 
\startsquarepar \indent In order to accurately capture the transition point from RR to MR and the Mach stem height, a mesh-independent study was performed. Different mesh sizes were implemented at the free-stream Mach number of 3, a reduced frequency $\kappa = 0.726$. The study was started with mesh 1, of $328 \times 90$ elements. Then, the mesh size was doubled to generate mesh 2, mesh 4, mesh 8, and mesh 16. The time step was allowed to change during each simulation, but it was controlled with an upper limit of the Courant–Friedrichs–Lewy (CFL) number, which was set to be $0.2$. Figure \ref{fig:fig4} indicates the variation of the Mach number along the mid-plane of symmetry ($y = 0.9$) at the sonic criterion ($M_{transition\ point} \cong 1$). It shows the convergence of the results with the increase in the mesh size. Moreover, the \stopsquarepar
\begin{table}[H]
\centering
\caption{Independent Mesh test: Absolute percentage error of the transition wave angle at the reflection point, $\beta_{t_{p}}$, and the Mach stem height, MS, at wedge angle, $\theta$=$27^\circ$.}
\begin{tabular}{ c c c c c c }
 \hline
 \multirow{2}{*}{Mesh \#} & \multirow{2}{*}{Mesh Size} & \multirow{2}{*}{$\beta_{t_{p}}\ (^{\circ})$} & \multirow{2}{*}{$\frac{\mathrm{MS}}{L(0)} \times 10^{-2}$} & \multicolumn{2}{c}{ $|Error|\%$}\\ & & & &
 $\beta_{t_{p}}$ & $\frac{\mathrm{MS}}{L(0)}$\\ [0.5ex]
 \hline
 1  & $328\ \times 90\ \ $& $41.39$ & 5.29 & $1.22$  &  46.4   \\ 
 2  & $656\ \times 180\ $ & $41.15$ & 7.34 & $0.62$  &  25.6  \\ 
 4  & $1312 \times 360\ $ & $41.03$ & 8.75 & $0.33$  &  11.3  \\
 8  & $2624 \times 720\ $ & $40.96$ & 9.53 & $0.17$  &  3.4   \\
 16 & $5248 \times 1440$  & $40.89$ & 9.87 &   -     &   -    \\[1ex]
 \hline
\end{tabular}
\label{table:2}
\end{table}
\noindent deviation between mesh 8 and mesh 16 in the Mach distribution can be neglected. Further, Table \ref{table:2} compares different mesh sizes, showing the absolute percentage error of the transition wave angle at the reflection point, $\beta_{t}$, and the Mach stem height, MS, at wedge angle, $\theta=27^\circ$. It is clear that for low mesh quality ($328 \times 90$), the error percentage of developing the Mach stem height, MS, at wedge angle of $\theta=27^{\circ}$, is around $50\%$, which is a huge error that would be a miss-leading in determining the dynamic transition point. Moreover, this error percentage rapidly declined by doubling the mesh size. Therefore, both mesh 8 and mesh 16 provide more precise results. In order to reduce the computational time, mesh 8 is selected to implement all the simulations. The minimum element size in mesh 8 is $0.75mm \times 0.78mm$, and the time step to ensure the CFL number of $0.2$ is $93.5 ns$.
\section{Verification}
\begin{figure}[H]
\centering
\includegraphics[width=0.99\linewidth]{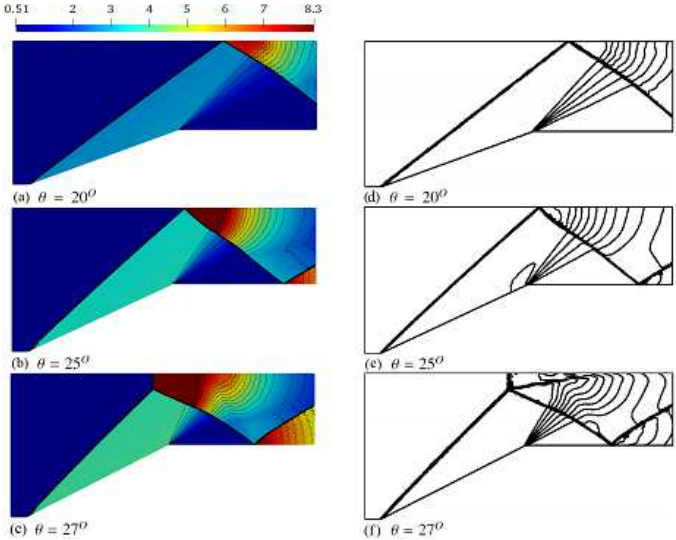}
\caption{Pressure contours at different wedge angles and $M_{t}=0.05$. The present work is shown in the colored left sub-figures, whereas Felthun and Skews [13] work is shown in the right sub-figures.}
\label{fig:fig6}
\end{figure}
\begin{figure}[H]
\centering
\includegraphics[width=0.99\linewidth]{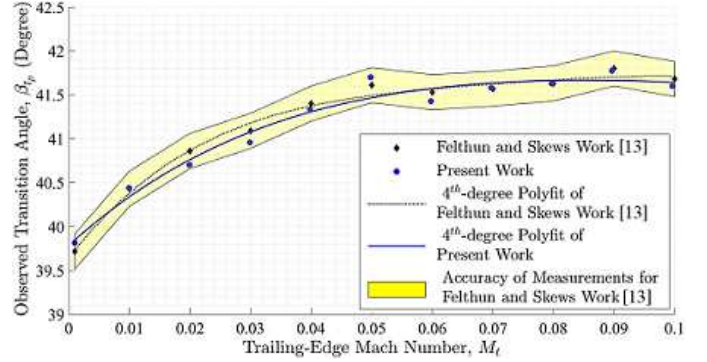}
\caption[Validation with the work of Felthun and Skews by measuring the effect of trailing-edge Mach number on transition angles from regular to Mach reflection]{Validation with the work of Felthun and Skews [13] by measuring the effect of trailing-edge Mach number on transition angles from regular to Mach reflection}
\label{fig:fig5}
\end{figure}

\indent The inserted dynamic code into the OpenFOAM solver is verified using the supersonic flow over a dynamic rotating wedge problem. The angle of a unity chord wedge is varied with different trailing-edge Mach numbers, $M_t$, in a flow of a free-stream Mach number of $3$. The observed transition angles, $\beta_t$, from regular to Mach reflection are obtained for these wedge rotation simulations. Figure \ref{fig:fig6} shows the comparison of the pressure contours for the flow over the rotating wedge with $M_t=0.05$ at different wedge angles, $\theta$. Further, Figure \ref{fig:fig5} indicates that our work is within the margin of accuracy of measurements for the work of Felthun and Skews \cite{felthun2004dynamic}.
\section{Results and Discussion}
The dynamic transition from RR to MR was studied using a horizontal moving wedge at a $M_{\infty}=3$ and was modeled by starting with RR in a steady flow with a wedge angle of $19^{\circ}$ and increasing it to $34^{\circ}$. The sonic and Detachment criteria were used to determine the transition point. Different reduced frequencies $(0.1:0.1:2)$ were tested to indicate the lag effect in the transition angles, $\theta_{t}$ and $\beta_{t}$, and the Mach stem height, MS. The used system properties and parameters are listed in Table \ref{table:1}.
\subsection{Dynamic Transition from RR to MR}
\begin{figure}[H]
    \centering
    \includegraphics[width=0.925\linewidth]{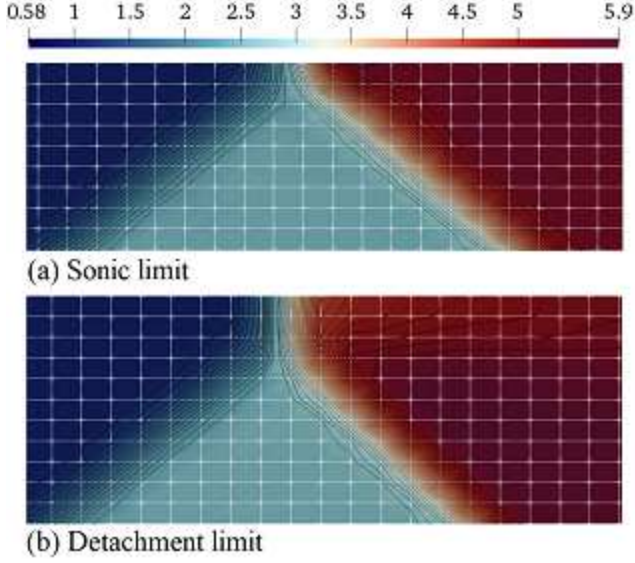}
    \caption{Density contours when the dynamic transition from RR to MR occurs with $\kappa = 0.1$ at: (a) Sonic and (b) Detachment criteria. }\label{fig:fig16}
\end{figure}

\begin{figure}[H]
    \centering
    \includegraphics[width=0.95\linewidth]{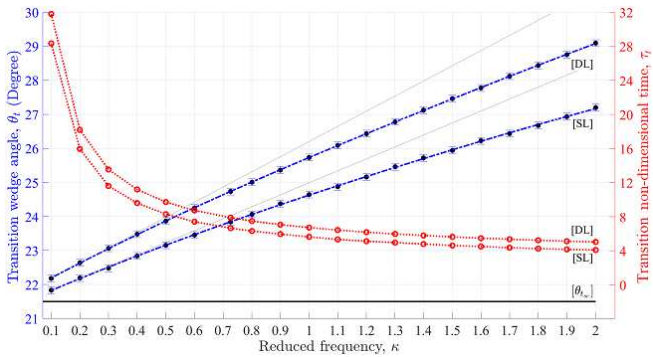}
    \caption{The variation of the transition wedge angle, $\theta_{t}$, and the transition non-dimensional time, $\tau_{t}$, with various reduced frequencies, $\kappa$ at the Sonic limit, SL, and the Detachment limit, DL. Gray straight lines are plotted to indicate the non-linear effect of the lag.}  \label{fig:fig8}
\end{figure}

\begin{figure}[H]
    \centering
    \includegraphics[width=0.95\linewidth]{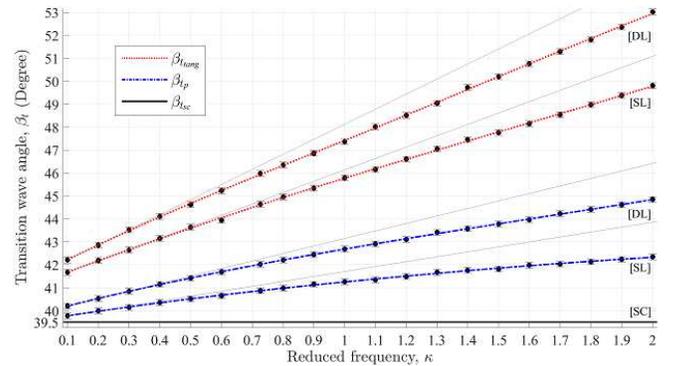}
    \caption{The variation of transition wave angle, $\beta_{t}$, with various reduced frequencies, $\kappa$ at the Sonic limit, SL, and the Detachment limit, DL. Gray straight lines are plotted to indicate the non-linear effect of the lag. }\label{fig:fig7}
\end{figure}

\begin{figure}[H]
    \centering
    \includegraphics[width=0.975\linewidth]{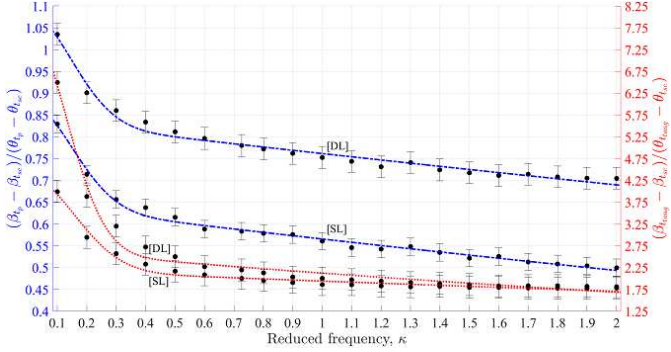}
    \caption{The variation of the lag in the transition wave angle to the lag in the transition wedge angle ratio, $(\beta_{t}-\beta_{t_{SC}})/(\theta_{t}-\theta_{t_{SC}})$, with various reduced frequencies, $\kappa$ at the Sonic limit, SL, and the Detachment limit, DL.}\label{fig:fig17}
\end{figure}

\indent The sudden motion of the wedge is started horizontally from the steady-state at $\theta=19^{\circ}$ with different reduced frequencies, $\kappa$. This movement affects the shock system. A lag appears in the incident shock wave angle behind the steady-state value at the same wedge angle, causing a curvature in the incident shock, which is indicated with the tangent wave angle, $\beta_{tang}$. Further, the strength of these effects depends on the value of $\kappa$. The dynamic transition from RR to MR is defined according to the sonic and detachment criteria. Figure \ref{fig:fig16} is a close view of two transition criteria over the used mesh at a $\kappa = 0.1$. The temporal time of the sonic criterion is determined when there is a point with a Mach number of $1$ on the top surface, as shown in Figure \ref{fig:fig16} (a). After a non-dimensional time of $3.46$ for the case of $\kappa=0.1$, a sharp bend in the reflected shock wave happens, detaching the shock from the symmetric surface and generating the triple point. This would decrease the Mach number below $1$ and when the Mach number reaches $0.475$ (Mach number after a normal shock) in a point on the symmetric plane, the detachment criterion is full-filled, as shown in Figure \ref{fig:fig16} (b). Further, in between the sonic and detachment limits, a Regular Reflection with Subsonic Downstream Flow (RRs) occurs instead of Regular Reflection (RR). In addition, Figure \ref{fig:fig12_13} shows the lag and curvature effects in the system with $\kappa = 0.5$, where the transition happens at a wedge angle at the sonic limit of $\theta_{t_{SL}}=23.15^{\circ}\pm 0.1^{\circ}$ and at the detachment limit at $\theta_{t_{DL}}=23.87^{\circ}\pm 0.1^{\circ}$, while the transition of the static case occurs at a wedge angle $\theta_{t_{SC}}=21.5^{\circ}$.
\startsquarepar \indent The investigation of the transition from RR to MR was conducted by studying the change of the transition non-dimensional times, $\tau_{t}$, transition wedge angles, $\theta_{t}$, and transition wave angles, $\beta_{t}$, at the two criteria with various $\kappa$. The increase in reduced frequency represents an increase in the velocity of the moving wedge. Thus, the required time to change the state will decrease with the increase of $\kappa$. After the significant decrease in the transition non-dimensional time, $\tau_{t}$ at low values of $\kappa$, it starts to gradually decrease at higher values of $\kappa$ than $0.4$, reaching $\tau_{t}=4$ from the starting wedge angle of $\theta=19^{\circ}$ at $\kappa=2$, as shown in Figure \ref{fig:fig8} right. Further, the change in the non-dimensional time \stopsquarepar 
\noindent between the detachment and sonic limits decreases, dramatically at low values of $\kappa$ and slightly at high values of $\kappa$, from $3.46$ in the case of $\kappa=0.1$ to $0.95$ in the case of $\kappa=2$. In addition, the lag in the flow appears with the wedge motion. Thus, there will be time needed to transfer the information from the wedge apex to the end of the incident shock. During this time, the wedge continues its motion. That is why the transition wedge angle in the dynamic cases is larger than that of the static case. This is shown in Figure \ref{fig:fig8} left with a Gauss curve fit and the uncertainty in measuring the angles is $\pm 0.1^{\circ}$. Consequently, increasing the motion rate of the wedge, $\kappa$, increases the gap between the dynamic and the static transition wedge angles. Hence, the reduced frequency generates a spatial and temporal lag in the system.\\
\indent The difference between the tangent transient wave angle, $\beta_{t_{tang}}$, and the wave angle at reflection$/$triple point, $\beta_{t_{p}}$, shown in Figure \ref{fig:fig7}, emphasizes the lag in the propagation of the information along the incident shock wave. Even at a very low value of $\kappa$, such as $\kappa=0.1$, the lag in $\beta_{t_{tang}}$ from the static case (about $2.5^{\circ} \pm 0.1^{\circ}$), which represents the curvature of the incident shock at the apex of the wedge, is relatively large with respect to the lag in $\beta_{t_{p}}$ (about $0.5^{\circ} \pm 0.1^{\circ}$). At the values of $\kappa$ smaller than $0.4$, the differences in wave angles, $\beta_{t_{p}}$ and $\beta_{t_{tang}}$, between the sonic and detachment criteria are within a fraction of a degree. Further, the differences increase gradually and reach $2.5^{\circ} \pm 0.1^{\circ}$ for $\beta_{t_{p}}$ and $3.2^{\circ} \pm 0.1^{\circ}$ for $\beta_{t_{tang}}$ at $\kappa = 2$.\\
\indent Another way to study the lag in the shock wave system is achieved with the ratio of the difference between the wave angles in the dynamic and static cases to the difference between the wedge angles in the dynamic and static cases, $(\beta_{t}-\beta_{t_{SC}})/(\theta_{t}-\theta_{t_{SC}})$, as shown in Figure \ref{fig:fig17}. This ratio represents the ratio between the speed of information propagation along the incident shock wave to the speed of wedge motion. The ratio decreases steeply at $\kappa$ smaller than 0.4; i.e. the speed of information propagation decreases dramatically with the increase of $\kappa$ at the low-value range. On the other hand, the ratio decreases almost linearly with a lower negative slope at the value of $\kappa$ larger than $0.4$.

\subsection{Mach Stem Height}
\begin{figure}[H]
    \centering
    \includegraphics[width=0.975\linewidth]{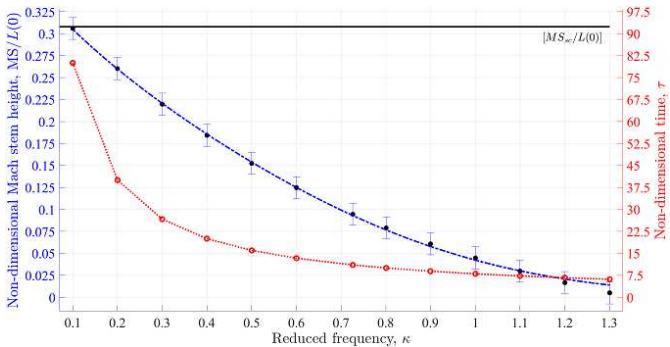}
    \caption{The variation of non-Dimensional Mach stem height and the corresponding Non-Dimensional time at wedge angle, $\theta=27^\circ$, with various reduced frequencies, $\kappa$.} \label{fig:fig9}
\end{figure}

In order to examine the effect of changing the movement rate of the wedge on the development of the Mach stem, MS, its height, $\textrm{MS}/L(0)$, was measured at a certain wedge angle, $\theta=27^{\circ}$, for different values of $\kappa$, as shown in Figure \ref{fig:fig10_11}. At the static case, a MR happens at wedge angle, $\theta_{SC}$, of $27^{\circ}$ with a $\textrm{MS}_{SC}/L(0) = 0.308$. On the other hand, the dynamic Mach stem height decreases with the increase of $\kappa$. At a very high value of $\kappa$, the $\textrm{MS}/L(0)$ disappears and the RR occurs, due to the excessive lag in transition, as shown in Figure \ref{fig:fig10_11}. Moreover, Figure \ref{fig:fig9} shows that longer time scales (small values of $\kappa$) enable the Mach stem to develop close to the static case value. For example, the $\textrm{MS}/L(0)$ is about $99.4\%$ of that of the static case at $\kappa=0.1$. Further, the deviation of the dynamic Mach stem height from the static case, $\textrm{MS}_{SC}/L(0) - \textrm{MS}/L(0)$, increases gradually with the increase of $\kappa$, reaching a percentage value of $98.4\%$ with respect to the static case at $\kappa=1.3$.
\subsection{Shock Reflection Domain}
\startsquarepar \indent Theoretically, when a supersonic flow impinges a tilted \stopsquarepar 
\begin{figure}[H]
    \centering
    \includegraphics[width=0.99\linewidth]{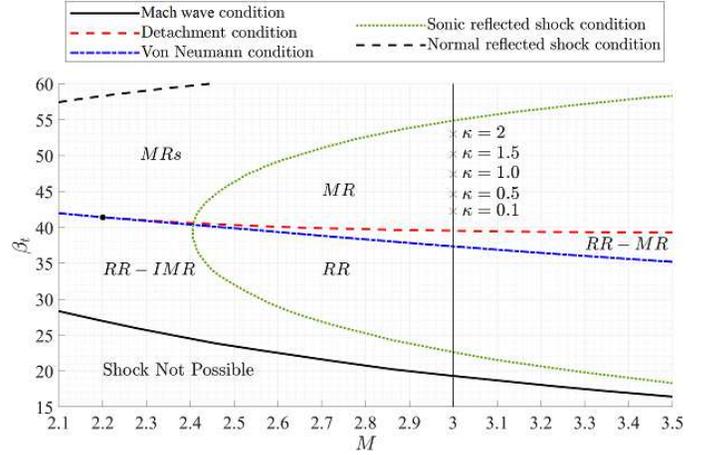}
    \caption{Comparison of dynamic transition wave angle, $\beta_{t_{tang}}$ with the analytical steady transition wave angle, $\beta_{t_{sc}}$.}\label{fig:fig14}
\end{figure}
\begin{figure}[H]
    \centering
    \includegraphics[width=0.99\linewidth]{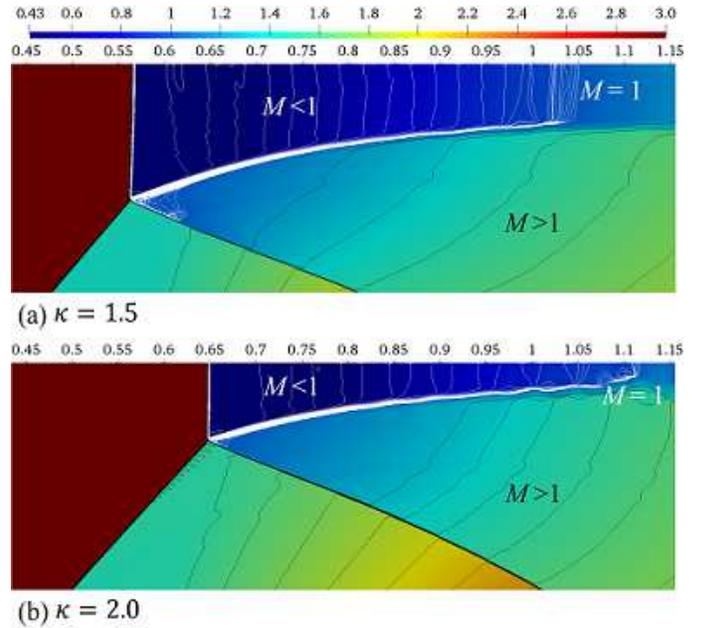}
    \caption{Mach contours at wedge angle of $\theta=34^\circ$, at different reduced frequencies, (a) $\kappa = 1.5$ and (b) $\kappa = 2.0$. The subsonic and supersonic regions are represented by white and black contours, respectively.}\label{fig:fig15}
\end{figure}
\noindent wedge, there are many possible shock wave systems, such as Regular Reflection (RR), Mach Reflection (MR), Mach Reflection with Subsonic Downstream Flow (MRs), Inverted Mach Reflection (IMR), etc, as mentioned by Mouton \cite{mouton2007transition}. The generation of a specific configuration depends on the free-stream Mach number and the wedge angle$/$wave angle. Figure \ref{fig:fig14} shows the physical limits of each shock configuration for $\gamma=1.4$, including the dynamic tangent transition wave angles, $\beta_{t_{tang}}$, at a free-stream Mach number of $3$ and different values of $\kappa$. The horizontal upstream motion of the wedge (increasing the wedge angle) generates a lag in $\beta_{t_{tang}}$, placing it above the DSD. Furthermore, increasing the reduced frequency moves the transition angle far away from the DSD. At high values of $\kappa$, $\beta_{t_{tang}}$ becomes very close to the sonic reflected shock condition (MRs). This was indicated by a small triangular subsonic zone in the reflection domain between the slip-line and the reflected shock at $\kappa=1.5$ and $2$ and $\theta=34^{\circ}$, as shown in Figure \ref{fig:fig15}. At $\kappa=2$, the triangular subsonic zone is very small due to the excessive lag in the shock system, which delays the development of MS. Thus, it became very close to the expansion fan that quickly accelerates the flow to supersonic speeds.
\section{Conclusion}
The current research work aims to investigate the dynamic transition from RR to MR over a two-dimensional slip wedge at a free-stream Mach number of 3. The trailing edge of the wedge moves horizontally upstream with various reduced frequencies, $\kappa=0.1:0.1:2$. The phenomenon was studied by analyzing the variation of the transition angles, and examining the lag in the development of the Mach stem with various reduced frequencies. Further, a comparison between the dynamic and theoretical static transition was achieved. The major conclusions of the study are:
\begin{itemize}[topsep=1pt,itemsep=1pt,partopsep=1pt, parsep=1pt]
\item The time scale needed to reach the transition declines rapidly at low values of $\kappa$, and slowly with high values of $\kappa$. However, the lag in the transition parameters, wedge angle, and wave angles, changes gradually with $\kappa$.
\item The ratio of the speed of information propagation along the incident wave to the speed of wedge motion falls quickly at lower values of $\kappa$ than $0.4$ and decreases with an almost constant lower rate at higher values.
\item Although the difference between the theoretical sonic and detachment conditions is within a degree, the difference gradually increases higher than a degree for $\kappa > 0.4$.
\item The growth of the Mach stem is very sensitive to the numerical resolution and the lag due to the motion of the wedge with a reduced frequency, $\kappa$.
\item At low values of $\kappa$, the dynamic transition occurs slightly above the theoretical detachment limit (MR). On the opposite side, the dynamic transition wave angle tends to reach the physical limit of the Mach Reflection with Subsonic Downstream Flow (MRs) at high values of $\kappa$.
\end{itemize}

\bibliography{Margha}
\end{multicols*}
\begin{figure*}
    \centering
    \includegraphics[width=0.99\linewidth]{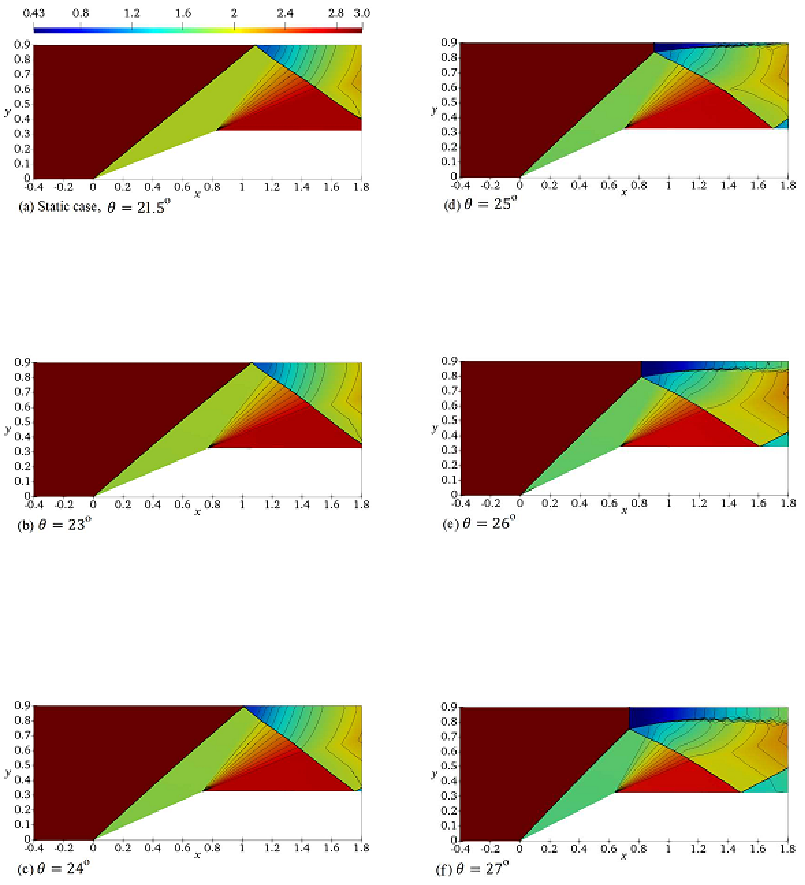}
    \caption{Mach contours at different wedge angles at $\kappa = 0.5$. The transition static case at $\theta=21.5^{\circ}$ is simulated to compare with the dynamic case.}\label{fig:fig12_13}
\end{figure*}

\begin{figure*}
    \centering
    \includegraphics[width=0.99\linewidth]{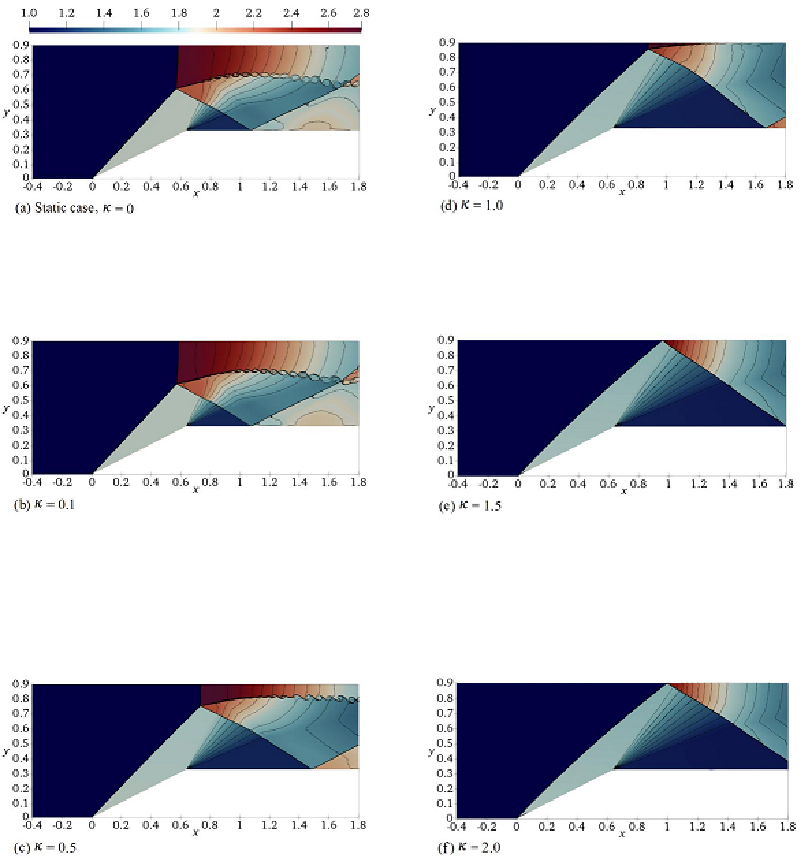}
    \caption{Temperature contours at wedge angle of $\theta = 27 ^\circ$, for the static case and different reduced frequencies, $\kappa$.}\label{fig:fig10_11}
\end{figure*}
\end{document}